\documentclass[conference]{IEEEtran}
\IEEEoverridecommandlockouts
\usepackage{cite}

\usepackage{amsmath,amssymb,amsfonts}
\usepackage{makecell}
\usepackage{graphicx}
\usepackage{textcomp}
\usepackage{amsmath}
\usepackage{booktabs} 
\usepackage{multirow}
\usepackage{hyperref}       
\usepackage{url}            
\usepackage{booktabs}       
\usepackage{amsfonts}       
\usepackage{nicefrac}       
\usepackage{microtype}      
\usepackage{xcolor}         
\usepackage{algorithm}
\usepackage{algpseudocode}
\usepackage{amsmath}
\usepackage{tikz}

\usepackage{orcidlink}
\usepackage{comment}
\usepackage{amssymb}
\usepackage{braket}
\usepackage{xcolor}
\usepackage{amsmath}
\usepackage{enumitem}

\setitemize{leftmargin=*}
\setenumerate{leftmargin=*}

\DeclareMathOperator*{\argmax}{arg\,max}

\def\BibTeX{{\rm B\kern-.05em{\sc i\kern-.025em b}\kern-.08em
    T\kern-.1667em\lower.7ex\hbox{E}\kern-.125emX}}
\definecolor{marineblue}{rgb}{0.05, 0.3, 0.57} 

\usepackage{titlesec}
\titlespacing\section{0pt}{0.6\baselineskip}{0.3\baselineskip}
\titlespacing\subsection{0pt}{0.3\baselineskip}{0.15\baselineskip}
\titlespacing\subsubsection{0pt}{0.2\baselineskip}{0.1\baselineskip}

\renewcommand{\IEEEbibitemsep}{0pt plus 2pt}
\makeatletter
\IEEEtriggercmd{\reset@font\normalfont\refsize}
\makeatother
\IEEEtriggeratref{1}

\begin{document}

\title{QADQN: Quantum Attention Deep Q-Network for Financial Market Prediction
\vspace{-10pt}
}

\author{\IEEEauthorblockN{Siddhant Dutta\textsuperscript{1}, Nouhaila Innan\textsuperscript{2,3}, Alberto Marchisio\textsuperscript{2,3}, Sadok Ben Yahia\textsuperscript{4}, and Muhammad Shafique\textsuperscript{2,3}
\IEEEauthorblockA{\textsuperscript{1}SVKM's Dwarakdas J. Sanghvi College of Engineering, India\\ 
\textsuperscript{2}eBRAIN Lab, Division of Engineering, New York University Abu Dhabi (NYUAD), Abu Dhabi, UAE\\
\textsuperscript{3}Center for Quantum and Topological Systems (CQTS), NYUAD Research Institute, NYUAD, Abu Dhabi, UAE\\
\textsuperscript{4}The Maersk Mc-Kinney Moller Institute, University of Southern Denmark, 
Alsion 2, 6400- Sønderborg, Denmark\\
siddhant.dutta180@svkmmumbai.onmicrosoft.com, nouhaila.innan@nyu.edu, alberto.marchisio@nyu.edu, say@mmmi.sdu.dk\\
muhammad.shafique@nyu.edu,\\
}}
\vspace{-30pt}
}

\maketitle

\begin{abstract}
Financial market prediction and optimal trading strategy development remain challenging due to market complexity and volatility. Our research in quantum finance and reinforcement learning for decision-making demonstrates the approach of quantum-classical hybrid algorithms to tackling real-world financial challenges. In this respect, we corroborate the concept with rigorous backtesting and validate the framework's performance under realistic market conditions, by including fixed transaction cost per trade. This paper introduces a Quantum Attention Deep Q-Network (QADQN) approach to address these challenges through quantum-enhanced reinforcement learning. Our QADQN architecture uses a variational quantum circuit inside a traditional deep Q-learning framework to take advantage of possible quantum advantages in decision-making. We gauge the QADQN agent's performance on historical data from major market indices, including the S\&P 500. We evaluate the agent's learning process by examining its reward accumulation and the effectiveness of its experience replay mechanism. Our empirical results demonstrate the QADQN's superior performance, achieving better risk-adjusted returns with Sortino ratios of 1.28 and 1.19 for non-overlapping and overlapping test periods respectively, indicating effective downside risk management.
 
\end{abstract}

\begin{IEEEkeywords}
Quantum Reinforcement Learning, Quantum Finance, Decision-Making
\end{IEEEkeywords}

\section{Introduction}
Financial markets are characterized by their complexity and unpredictability, driven by many factors that fuel volatility and challenge the efficacy of predictive models. Traditional methods in computational finance often struggle to fully utilize the vast datasets available due to limitations in computational power and algorithm complexity \cite{app9245574}. The complexity arises from various interacting elements such as macroeconomic indicators, market sentiment, and geopolitical events contributing to market volatility \cite{Chou2015,NIU2023102738,10.1093/jjfinec/nbaa024}.

Volatility in financial markets poses significant challenges for both prediction and risk management. Traditional volatility models, like GARCH and its variants, have been widely used but often fall short of capturing market behaviors' intricate and dynamic nature. The limitations of these models highlight the need for more advanced approaches that can handle non-linearities and high-dimensional data.

Financial statement data, alongside information extracted from business news, can be integrated with advanced ML algorithms to generate investment signals or predict a company's future performance \cite{10.1145/3383455.3422540}. This approach enhances the stock screening process by identifying promising investment opportunities. However, while these algorithms effectively address stock selection, they do not inherently resolve the issue of optimal position sizing and allocation among the chosen investments. Consequently, the trader must still exercise discretion in determining the precise timing for entering and exiting positions \cite{Zhang_Skiena_2010}. 

Current deep learning methodologies, such as Recurrent Neural Networks (RNNs), Long Short-Term Memory Networks (LSTMs), and Transformer models, have been employed to predict stock prices by learning from historical data. These models excel at capturing temporal dependencies and patterns in the data, leading to improved predictive performance. However, they often require extensive computational resources and can be prone to overfitting, especially when dealing with noisy and non-stationary financial data \cite{asi6060106,zou2023stockmarketpredictiondeep}.

In addition to employing Machine Learning (ML) and deep learning techniques, traditional stochastic process models such as Geometric Brownian Motion (GBM) offer a reliable method for stock price prediction by simulating market returns as a continuous-time random walk. The GBM model posits that the logarithm of stock prices follows a normal distribution, thereby incorporating the inherent randomness and volatility of financial markets. However, this assumption also presents a limitation, as it may not fully capture extreme market events or non-linear dynamics \cite{e22121432,LADDE2009e1203}. The significance of GBM extends to its foundational role in the development of the Black-Scholes option pricing model, which is extensively applied in the valuation of financial derivatives \cite{quayesam2024modelingstockpricedynamics}. This model's simplicity and interpretability make it a valuable tool for stock price prediction, provided historical price data is available \cite{zamani2022collectivebehaviorstockprices}. 

The QADQN framework addresses the limitations of computational overhead in attention mechanisms by replacing them with a Multi-Head Quantum Attention layer. It also improves explainability by integrating quantum attention mechanisms, utilizing the computational power of quantum computing to enhance the processing and interpretation of large-scale financial data. Using imitative learning strategies helps to solve the problem of finding a balance between exploitation and exploration \cite{belkhale2024data}. The trading agent is instructed via the use of a Q-learning algorithm and a replay buffer that is initially filled with actions taken from the Dual Thrust approach \cite{chen2021reinforcement}. These methods provide the network with enhanced expertise in the financial area when they are integrated into the POMDP framework. 

\textbf{The novel contributions of this work can be summarized as follows:}

\begin{itemize}
    \item Incorporation of \textit{quantum attention layers} within deep Reinforcement Learning (RL) agents to enhance the framework's ability to focus on relevant market features, resulting in improved decision-making efficiency.
    
    \item Application of the QADQN within a \textit{Partially Observable Markov Decision Process (POMDP) framework} utilizing attention-based deterministic policy gradient method, which is well-suited for dealing with the uncertainty and partial observability inherent in financial markets.
    
    \item Through modified Q-learning and the initialization of a replay buffer, pre-populated with actions derived from the established Dual Thrust methodology, imitative learning approaches in quantum agents aim to maintain an optimal balance between exploration and exploitation of available resources.
    
    \item Backtesting and validation against historical market data, including major indices like the S\&P 500. The backtesting process includes considerations for transaction costs and estimates the total return rate, Maximum Drawdown, and the final Sharpe and Sortino ratio, providing an assessment of the framework's reliability.
    
    \item Development of an attention deterministic policy gradient method tailored to handle the POMDP setting for quantum agents. This method enhances the framework's ability to learn effective trading strategies and trends, thereby improving its long-term reward accumulation.
    
\end{itemize}
Sec. \ref{s2} reviews Quantum Computing (QC) and RL in financial markets; Sec. \ref{s3} explains the QADQN, POMDP, and dual thrust strategy; Sec. \ref{s4} compares QADQN's performance with prior methods and different market scenarios; Sec. \ref{s5} concludes with key findings and future directions for quantum-enhanced RL in trading.


\begin{figure*}[!ht]
    \noindent
    \includegraphics[width=\linewidth]{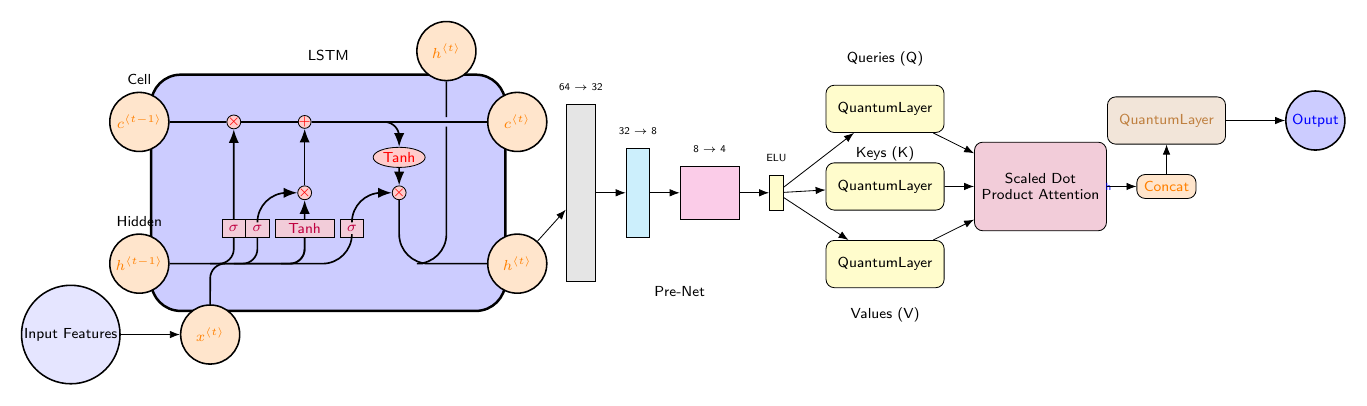}
    \vspace{-1.cm}
     \caption{QADQN Agent Architecture: The architecture begins with an LSTM network, which processes the input features along with the previous hidden and cell states \((h(t-1), c(t-1))\) to produce the current hidden and cell states \((h(t), c(t))\). The output of the LSTM is then passed through a pre-net consisting of linear layers reducing the dimensionality from 64 to 32, then to 8, and finally to 4, which is the number of qubits. These outputs are fed into quantum layers to generate the Queries (Q), Keys (K), and Values (V) for the scaled dot-product attention mechanism. The attention mechanism integrates these quantum-derived components, followed by concatenation and an additional quantum layer to output Q-values, which are used for action selection in OHLC data processing.}
    \label{fig:QADQN}
\end{figure*}

\section{Background and Related Work \label{s2}}

\subsection{Quantum Computing for Finance}

Recent advancements in QC have demonstrated significant potential for financial market applications, particularly in enhancing prediction accuracy and decision-making efficiency. Quantum Machine Learning (QML) emerges as a promising field \cite{biamonte2017quantum,zaman2023survey}, combining the computational power of quantum systems with ML algorithms to analyze financial data more effectively \cite{innan2024financial,innan2024financial1}.

Consider a quantum state representing a financial portfolio:
\begin{equation}
    \ket{\psi_\text{portfolio}} = \sum_i \alpha_i \ket{a_i},
\end{equation}
where $\ket{a_i}$ represents individual assets and $\alpha_i$ their corresponding weights \cite{herman2022survey}, subject to the normalization condition $\sum_i |\alpha_i|^2 = 1$.

Portfolio optimization can be formulated as a Quadratic Unconstrained Binary Optimization (QUBO) problem:
\begin{equation}
    \min f(x) = x^T Q x + c^T x,
\end{equation}
where $x$ is a binary vector representing asset selection, $Q$ is the covariance matrix, and $c$ represents expected returns. This can be mapped to a quantum Hamiltonian:
\begin{equation}
    H = \sum_{ij} Q_{ij} \sigma^z_i \sigma^z_j - \sum_i c_i \sigma^z_i,
\end{equation}
where $\sigma^z_i$ are Pauli-Z operators. Quantum algorithms like Quantum Approximate Optimization Algorithm (QAOA) can be applied to find the ground state of this Hamiltonian, representing the optimal portfolio allocation:
\begin{equation}
    \ket{\psi_\text{QAOA}} = \prod_p e^{-i\beta_p H_B} e^{-i\gamma_p H_C} \ket{\psi_\text{init}},
\end{equation}
where $H_B$ and $H_C$ are mixing and cost Hamiltonians respectively, and $\beta_p$, $\gamma_p$ are variational parameters \cite{farhi2014quantum}.

For risk assessment, quantum entanglement can model complex correlations. Value-at-Risk (VaR) is a widely used measure in risk management to quantify the potential loss on an asset or portfolio over a specific time horizon and with a certain confidence level:
\begin{equation}
    \text{VaR}_{\alpha}[L] = \inf \{ p \in \mathbb{R}^+ : \langle \psi | \hat{P}_L \leq p | \psi \rangle \geq \alpha \},
\end{equation}
where \(\hat{P}_L \leq p\) is a projection operator onto the subspace where the loss is less than or equal to \(p\).

Conditional Value-at-Risk (CVaR) \cite{egger2020credit}, or expected shortfall, is the expected loss given that the loss exceeds the VaR threshold. It can be expressed as:
\begin{equation}
    |\psi_{\text{VaR}}\rangle = \frac{\hat{P}_{L > \text{VaR}_{\alpha}} |\psi\rangle}{\sqrt{\langle \psi | \hat{P}_{L > \text{VaR}_{\alpha}} | \psi \rangle}},
\end{equation}
\begin{equation}
    \text{CVaR}_{\alpha}[L] = \langle \psi_{\text{VaR}} | \hat{L} | \psi_{\text{VaR}} \rangle.
\end{equation}
The Economic Capital Requirement (ECR) is defined as the difference between VaR and the expected loss:
\begin{equation}
    \text{ECR}_{\alpha}[L] = \text{VaR}_{\alpha}[L] - \langle \psi | \hat{L} | \psi \rangle.
\end{equation}
\subsection{Quantum Machine Learning for Finance}

QML algorithms, such as quantum neural networks, can be used for financial time series prediction \cite{10.1007/978-3-031-19493-1_6,venturelli2019reverse}. These quantum algorithms leverage superposition and entanglement to process large financial datasets more efficiently. One promising field within QML is Quantum Reinforcement Learning (QRL) \cite{dong2008quantum}, which integrates QC with RL to develop advanced trading strategies. In QRL, quantum circuits are used to enhance exploration and exploitation mechanisms within RL frameworks. The basic RL model can be represented as:
\begin{equation}
Q^{*}(s, a) = \max_{\pi} \mathbb{E}_{\pi} \left[ \sum_{t=0}^{\infty} \gamma^t r_{t+1} \mid S_t = s, a_t = a \right],
\end{equation}
where \( Q^{*}(s, a) \) is the optimal action-value function, \( \gamma \) is the discount factor, \( r_t \) is the reward at time \( t \), and \( \pi \) is the policy.

However, integrating quantum circuits into RL frameworks poses significant challenges. For instance, Quantum Deep Q-Networks (QDQN) and Quantum Policy Gradient methods have shown theoretical benefits, but issues like quantum noise, decoherence, and scalability impede practical implementation \cite{franz2023uncovering}. Furthermore, the optimization of VQC is represented as:
\begin{equation}
\theta^* = \arg\min_{\theta} \langle \psi(\theta) \mid H \mid \psi(\theta) \rangle,
\end{equation}
and it requires careful tuning of quantum parameters \( \theta \), which can be computationally intensive in terms of convergence. Recent advancements in quantum attention mechanisms have demonstrated potential in enhancing the performance of ML, including computer vision and high energy physics tasks \cite{10.1007/s10489-024-05337-w,axioms13030187,zhao2022qsan}. These mechanisms leverage QC principles to efficiently handle the high-dimensional parameter spaces associated with self-attention, a crucial component in many contemporary deep-learning architectures \cite{10.5555/3295222.3295349}.

\section{QADQN Framework\label{s3}}

The Quantum Attention Deep Q-Network (QADQN) framework extends the concept of hybrid quantum-classical architectures to the domain of RL. By integrating NISQ-based quantum self-attention mechanisms with deep Q-networks \cite{shi2023natural, chen2020variational,lockwood2020reinforcement,skolik2022quantum}, QADQN aims to improve decision-making processes in complex environments. Our approach involves carefully optimizing Variational Quantum Circuit (VQC) parameters using the backpropagation method for gradient computation, addressing convergence challenges in high-dimensional quantum-classical hybrid systems in the financial domain.

The input to the QADQN is an $n$-day state representation of financial data transformed using a logarithmic function, denoted as $S_t \in \mathbb{R}^{n \times f}$, where $n$ is the number of days, and $f$ is the number of features per day. Each row of the state $S_t$ at time $t$ is given by:
\begin{equation}
S_{t,i} = \left[\ln(p_{t-i+1}/p_{t-i}), \ldots, \ln(f^f_{t-i+1}/f^f_{t-i})\right]
\end{equation}
where $p_t$ represents the close price at time $t$, $f^j_t$ represents the $j$-th open-high-low-close (OHLC)-based feature at time $t$, and $i$ ranges from 1 to $n$, covering the $n$-day window.
\subsection{Pre-Neural Net}
The LSTM layer processes the input sequence and outputs a sequence of hidden states $h_t \in \mathbb{R}^{64}$ for each time step. For single-step predictions, we use the last hidden state, while for sequence outputs, we use the full sequence of hidden states\cite{zou2024novel}, as shown in Fig.~\ref{fig:QADQN}:
\begin{equation}
h_t, c_t = \text{LSTM}(S_t, h_{t-1}, c_{t-1}),
\end{equation}
where $h_t$ is the hidden state and $c_t$ is the cell state at time $t$.

\begin{equation}
x_t = \text{ELU}(W_L \cdot \text{ELU}(W_{L-1} \cdot \ldots \text{ELU}(W_1 \cdot h_t + b_1) + b_{L-1}) + b_L).
\end{equation}

While attention mechanisms, similar to those in transformer models, can be employed independently, our hybrid approach enables the capture of long-term dependencies by extracting temporal features with LSTM and refining the LSTM's contextualized output with the quantum attention mechanism.

\subsection{Quantum Attention Mechanism}
The core of our framework is the Quantum Multihead Self-Attention (QMSA) mechanism. This mechanism projects the input state to a quantum state using parameterized quantum circuits, computes attention scores, and aggregates the results:
\paragraph{Data Encoding} Each input row $x_i$ of the LSTM output is encoded into a quantum state using a data loader operator $U^\dagger(x_i)$:
\begin{equation}
|x_i\rangle \equiv U^\dagger(x_i)|0\rangle = \bigotimes_{j=1}^{dh} R_x^\dagger(x_{ij})|0\rangle,
\end{equation}
where $R_x$ is a parameterized rotation around the x-axis and \( dh \) represents the dimension of the hidden state.
\paragraph{Key and Query Operations} For each input row $x_i$, key operator $K^\dagger(\theta_K)$ and query operator $Q^\dagger(\theta_Q)$ are applied:
\begin{equation}
K_i = \langle x_i|K^\dagger(\theta_K)Z_0K(\theta_K)|x_i\rangle,
\end{equation}
\begin{equation}
Q_i = \langle x_i|Q^\dagger(\theta_Q)Z_0Q(\theta_Q)|x_i\rangle,
\end{equation}
where $Z_0$ represents a spin measurement of the qubit in the z-direction.
\paragraph{Value Operation} Each row of the input is passed through a value operator $V^\dagger(\theta_V)$ to obtain the value matrix $V$:
\begin{equation}
V_{ij} = \langle x_i|V^\dagger(\theta_V)Z_jV(\theta_V)|x_i\rangle.
\end{equation}
The circuit used to create $K$, $Q$, and $V$ is shown in Figure \ref{fig:QuantumLayer}.
\paragraph{Hybrid Attention Head Computation} The attention matrix \( A \) is computed using key and query vectors as \( A_{ij} = -(Q_i - K_j)^2 \). The final output is determined by applying softmax to \( A \) normalized by \( \sqrt{dh} \), then multiplying the result by the value matrix \( V \), yielding \( \text{SoftMax}\left(\frac{A}{\sqrt{dh}}\right) \cdot V \). 

\subsection{Quantum Post-Net Layer}
\begin{figure}
    \centering
    \includegraphics[height=3.5cm, width=\linewidth]{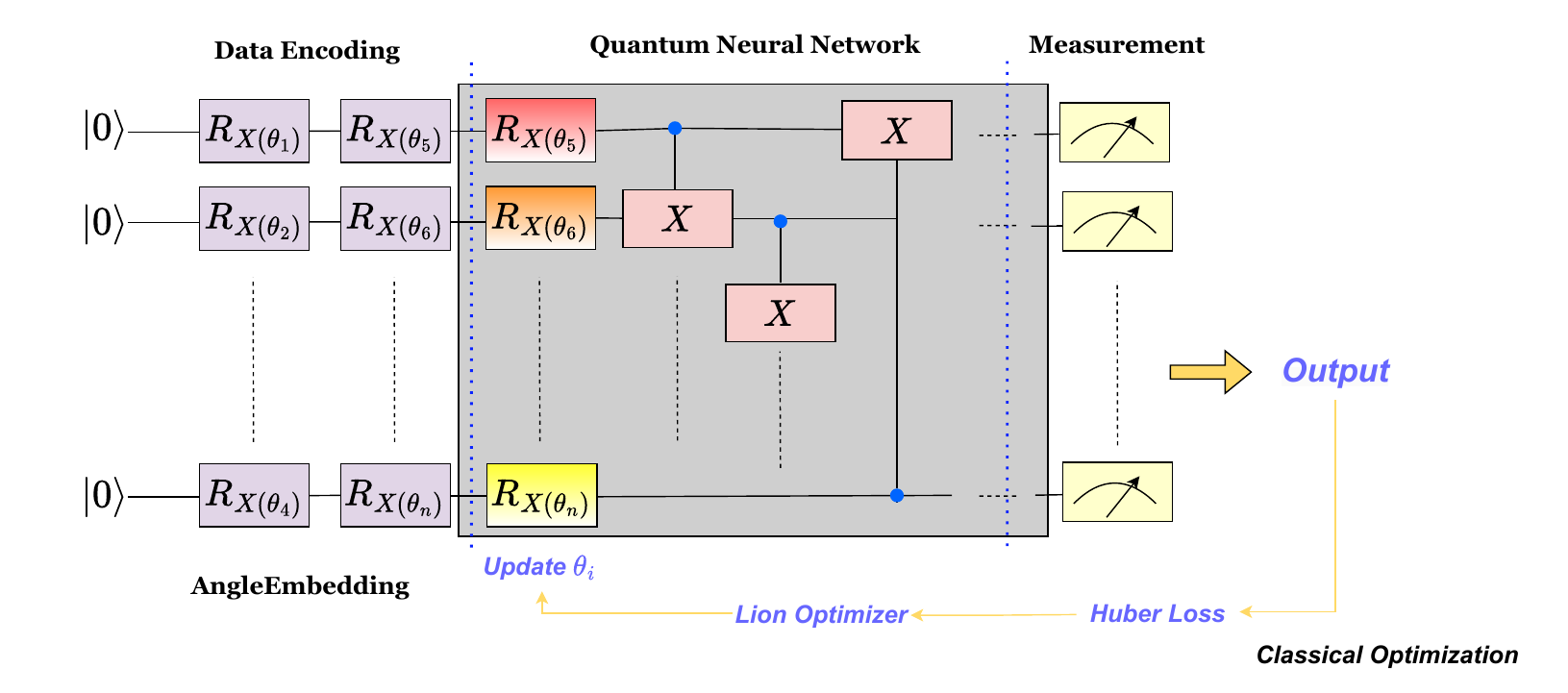}
    \vspace{-0.8cm}
     \caption{Quantum layer circuit: It begins with data encoding using angle embedding, where input data is encoded into the quantum state via parameterized rotation gates \(R_X(\theta_i)\). The encoded quantum states are then processed by a parameterized quantum circuit consisting of additional parameterized rotation gates and Controlled-NOT gates, facilitating entanglement and complex quantum state manipulation. The quantum states are subsequently measured to obtain classical outputs. During training, the classical outputs compute the Huber loss, and the parameters \(\theta_i\) are updated using the Lion optimizer. }
    \label{fig:QuantumLayer}
\end{figure}

The VQC is defined using the PennyLane framework. The circuit consists of an angle embedding layer followed by entangling layers as shown in Fig. \ref{fig:QuantumLayer}:
\begin{equation}
\ket{\psi(\theta)} = M(\theta) \ket{0}^{\otimes n},
\end{equation}
where $M(\theta)$ represents the parameterized quantum circuit and $n$ is the number of qubits. The circuit is defined as follows:
\begin{equation}
M(\theta) = \prod_{l=1}^{L} \left( \prod_{i=1}^{n} R_X(\theta_{l,i}) \right) \left( \prod_{i=1}^{n-1} \text{CNOT}(i, i+1) \right),
\end{equation}
where $L$ is the number of layers, $R_X(\theta)$ is the X-rotation gate, and $\text{CNOT}(i, j)$ is the controlled-NOT gate with qubit $i$ as control and qubit $j$ as target.
The expectation values of Pauli-Z observables are measured, which act as the final Q-Values:
\begin{equation}
q_t = (\langle Z_1 \rangle, \ldots, \langle Z_n \rangle).
\end{equation}

\vspace{-20pt}

\subsection{Agent}
The QADQN agent is designed to make trading decisions based on historical financial data. The agent's architecture consists of several key components:

\begin{itemize}    
    \item Action space: $\mathcal{A} = \{sit, buy, sell\}$.
    
    \item Policy network ($Q$): A quantum-classical hybrid network for action selection.
    
    \item Target network ($\hat{Q}$): A periodically updated copy of the policy network for stable learning.
    
    \item Prioritized Replay memory: For experience replay and prioritized experience replay.
\end{itemize}

The QADQN agent employs an Upper Confidence Bound (UCB) policy for action selection \cite{UCBZHang}, and integrates the Dual Thrust pre-populated actions from prioritized experience replay memory for the final action decision:

Action Selection Policy (UCB):
\begin{equation}
P(a|s) = \begin{cases}
1 - c + c \frac{\log(t)}{N_t(s)}, & \text{if } a = \arg\max_a Q(s, a) \\
\frac{c \log(t)}{N_t(s)}, & \text{otherwise}
\end{cases},
\end{equation}
where \( c \) is a constant determining the exploration-exploitation trade-off, \( t \) is the total number of time steps, and \( N_t(s) \) is the number of times state \( s \) has been visited up to time \( t \).

Dual Thrust Strategy:
\begin{equation}
\begin{aligned}
\text{Range} &= \max[\text{HH} - \text{LC}, \text{HC} - \text{LL}], \\
\text{BuyLine} &= \text{Open} + K1 \times \text{Range}, \\
\text{SellLine} &= \text{Open} - K2 \times \text{Range},
\end{aligned}
\end{equation}
where \( \text{Open} \) represents the day's opening price, \( K1 \) and \( K2 \) are constants controlling market resistance levels against breaking BuyLine and SellLine, respectively. \( \text{HH} \), \( \text{LC} \), \( \text{HC} \), and \( \text{LL} \) denote the highest high and lowest close prices over the previous periods \cite{cheng2024mot}.

\subsection{Training Algorithm}
The QADQN agent is trained using a combination of experience replay and prioritized experience replay. The training process is detailed in Algorithm \ref{alg:qadqn}.
\begin{algorithm}
\caption{QADQN Training Algorithm}
\label{alg:qadqn}
\begin{algorithmic}[1]
\State Initialize replay memory $\mathcal{D}$ to capacity $N$
\State Initialize prioritized replay memory $\mathcal{PD}$ to capacity $M$
\State Initialize action-value function $Q$ with random weights $\theta$
\State Initialize target action-value function $\hat{Q}$ with weights $\theta^- = \theta$
\For{episode = 1 to E}
    \State Initialize state $S_1$
    \For{t = 1 to T}
        \State Compute UCB action values $UCB(S_t, a)$ for all actions $a$
        \State Select action $a_t = \argmax_a UCB(S_t, a)$
        \State Execute action $a_t$ and observe reward $r_t$ and next state $S_{t+1}$
        \State Store transition $(S_t, a_t, r_t, S_{t+1})$ in $\mathcal{D}$
        \State Sample transitions from $\mathcal{D}$ according to priorities
        \State Compute priorities for sampled transitions
        \State Perform gradient descent step w.r.t. $\theta$ on $(y_j - Q(S_j, a_j; \theta))^2$
        \State Update priorities in $\mathcal{PD}$ based on TD error
        \If{t mod target\_update\_frequency == 0}
            \State $\theta^- \gets \theta$
        \EndIf
    \EndFor
\EndFor
\end{algorithmic}
\end{algorithm}
\begin{figure*}[ht]
    \centering
    \begin{tikzpicture}
        \node[anchor=south west,inner sep=0] (image) at (0,0) {\includegraphics[width=0.7\linewidth]{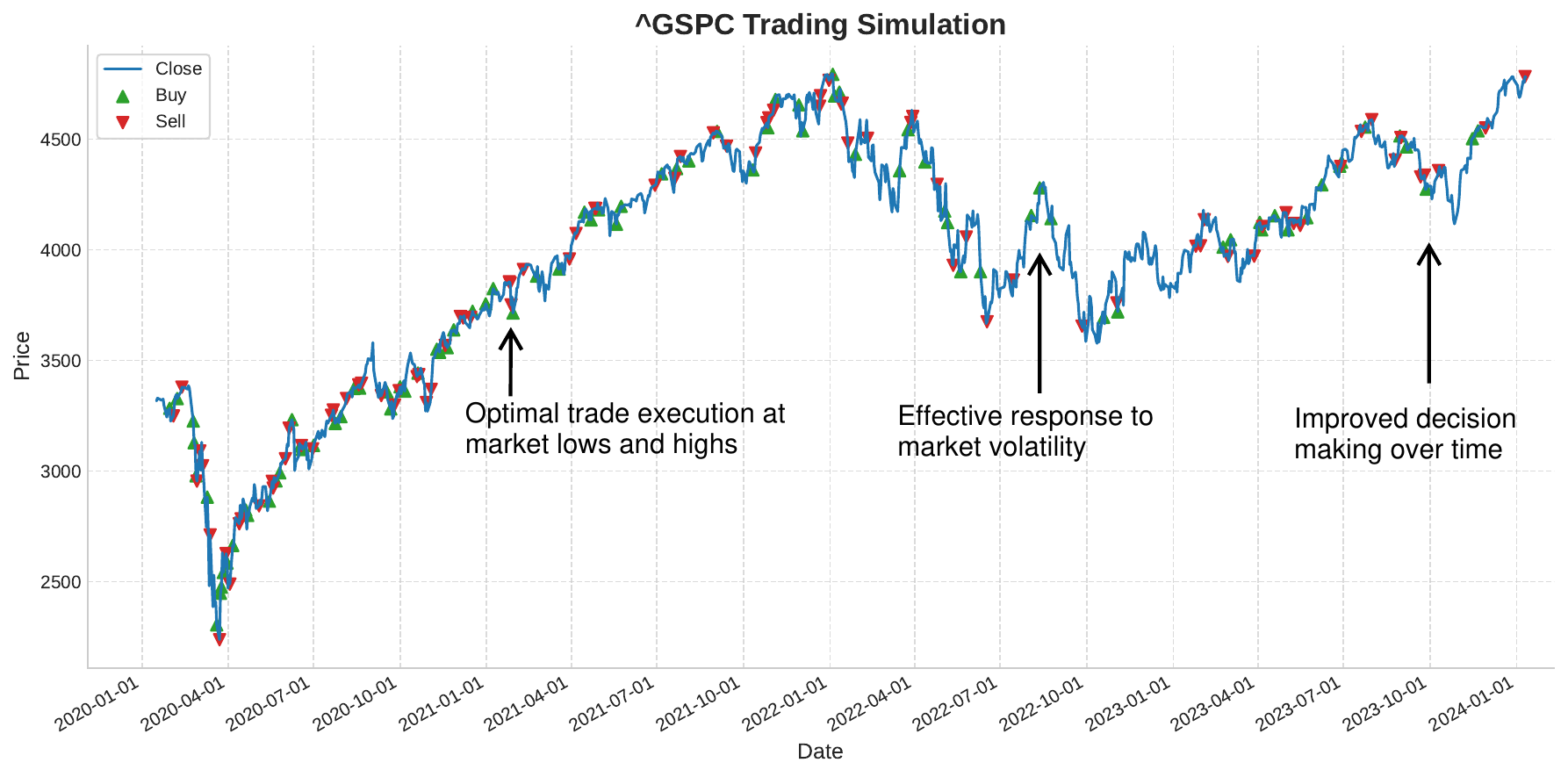}};
        
        \begin{scope}[x={(image.south east)},y={(image.north west)}]
            \node[align=left, text width=0.3\linewidth, font=\scriptsize, right,text=marineblue] at (1.0, 0.5) {
                \textbf{Key Observations:}
                \begin{itemize}
                    \item \textbf{Optimal Trade Execution:} Strategic buy (green triangles) and sell (red triangles) points effectively capitalize on market lows and highs.
                    \item \textbf{Market Volatility Response:} The framework effectively responds to significant market fluctuations, particularly noticeable during mid-2020 and early 2022.
                    \item \textbf{Progressive Decision Making:} There is an improvement in decision-making over time, suggesting that QADQN adapts or optimizes based on past performance and emerging trends.

                \end{itemize}
            };
        \end{scope}
    \end{tikzpicture}
    \vspace{-0.6cm}
    \caption{QADQN Agent trading simulation for S\&P 500 showcasing trading performance over time.}
    \label{fig:trade}
    \end{figure*}

Lion (EvoLved Sign Momentum) optimizer \cite{chen2024symbolic} is used for the classical optimization step. It only tracks momentum and uses the sign operation to calculate updates, leading to lower memory overhead and uniform update magnitudes across all dimensions. Lion has shown improvements in training various models for different tasks, especially in networks with Attention layers, compared to optimizers like Adam and Adafactor. It is expressed as:
\begin{equation}
\theta_t = \theta_{t-1} - \alpha \cdot \text{sign}(m_t) - \alpha \lambda \theta_{t-1},
\label{23}
\end{equation}
where $\theta_t$ is the parameter at time step t, $m_t$ is the exponential moving average of the gradient, $\alpha$ is the learning rate, $\lambda$ is the weight decay coefficient and $\text{sign}(x)$ is the sign function. Due to the usage of sign operation as seen in equation \ref{23}, it balances the gradient history and current gradient weighting, leading to better global optimization. This makes the gradients smooth and continuous, leading to better decision-making by the QADQN. The loss function used is Huber loss as it proves advantageous in scenarios where observed rewards occasionally contain outliers—instances where unrealistically large negative or positive rewards may occur during training but not during testing, and it can be described as:
\begin{equation}
    L = H(Q(S_t, a_t; \theta) - (r_t + \gamma \max_{a'} \hat{Q}(S_{t+1}, a'; \theta^-))),
\end{equation}
where $H$ is the Huber loss function. Unlike L2 loss, which could be significantly affected by such outliers, Huber loss maintains robustness by moderating the impact of these anomalies on the training process \cite{wang2019improved, 10.1007/s00521-022-07606-6}

\section{Results and Discussion\label{s4}}
\subsection{Experimental Setup}
The experiments are conducted using the S\&P 500 as the standard dataset with OHLC fetched from \textit{Yahoo finance API} \cite{yahoo_finance_api}, with a training period from January 15, 2016, to January 15, 2020, and evaluation on two test periods: an overlapping period from January 16, 2019, to January 16, 2024, and a non-overlapping period from January 16, 2020, to January 16, 2024. The QADQN framework is implemented with 4 qubits for each quantum layer, a 24-day window size, a discount factor ($\gamma$) of 0.95, and 200 training episodes. The Dual Thrust strategy parameters are set to $k1 = 0.8$ and $k2 = 0.4$. The framework is trained using both standard experience replay and prioritized experience replay.

\subsection{Performance Evaluation}

The QADQN model's performance is evaluated using both visual analysis of trading actions and quantitative metrics derived from backtesting and is compared to the Buy \& Hold Method and Deep Deterministic Policy Gradient (DDPG), which is an off-policy A2C model \cite{mironowicz2024applications}. 

\subsubsection{Trading Visualization}
Fig. \ref{fig:trade} presents a visual representation of the model's trading activities superimposed on the S\&P 500 price chart for the non-overlapping test period. The green markers indicate buy actions, predominantly occurring at lower price points of the index, while the red markers denote sell actions, typically executed at higher price levels. This pattern suggests that the agent has developed a strategy that potentially capitalizes on price fluctuations, buying at relative lows and selling at relative highs. Such behavior indicates the model's ability to identify and exploit short-term price trends within the market.

\subsubsection{Backtesting Results}

Backtesting is performed using the Backtest library with a commission rate of 0.2\%. The results for both test periods are summarized in Table 
\ref{table:backtesting}.
\begin{table}[h]
\caption{Backtesting results and Comparison for QADQN framework for S\&P 500.}
\label{table:backtesting}
\centering
\begin{tabular}{llll}
\toprule
\makecell{} & Metric & \makecell{Overlapping \\ Period} & \makecell{Non-Overlapping \\ Period} \\
\midrule
\multirow{4}{*}{\makecell{{\textbf{QADQN}}}} & Return [\%] & \textbf{78.91} & \textbf{71.26} \\
 & Sharpe Ratio & \textbf{0.82} & \textbf{0.77} \\
 & Sortino Ratio & \textbf{1.19} & \textbf{1.28} \\
 & Max. Drawdown [\%] & 19.24 & 19.24 \\
\midrule
\multirow{3}{*}{\makecell{\textbf{Buy \& Hold}}} & Return [\%] & 74.42 & 44.22 \\
 & Sharpe Ratio & 0.67 & 0.52 \\
 & Sortino Ratio & 1.07 & 0.85 \\
\midrule
\multirow{3}{*}{\makecell{{\textbf{DDPG}}}} & Return [\%] & 31.23 & 15.94 \\
 & Sharpe Ratio & 0.42 & 0.36 \\
 & Sortino Ratio & 0.81 & 0.79 \\

\bottomrule
\end{tabular}
\end{table}

\subsection{Discussion}

The QADQN framework demonstrates strong potential in applying quantum-enhanced RL to financial trading. The framework not only achieves substantial positive returns but also outperforms the Buy \& Hold and DDPG benchmarks in both test periods, particularly in the non-overlapping period.

\textit{Total Return:} The QADQN framework achieves significant positive returns in both test periods, with 75.37\% for the overlapping period and 66.82\% for the non-overlapping period. Notably, the framework outperforms the buy-and-hold strategy in both cases, with a particularly substantial outperformance in the non-overlapping period.

\textit{Risk-Adjusted Performance:} The Sharpe ratios of 0.78 and 0.75 for the overlapping and non-overlapping periods, respectively, indicate good risk-adjusted returns. The higher Sortino ratios (1.14 and 1.33) suggest that the framework manages downside risk very effectively, especially in the non-overlapping period, suggesting good generalization to unseen market conditions.

\textit{Maximum Drawdown:} Both test periods exhibit a maximum drawdown of 19.24\%, indicating the framework's ability to limit significant losses during market downturns.

\section{Conclusion\label{s5}}
This paper introduces a QADQN, which integrates a VQC within a traditional deep Q-learning network, aimed at emphasizing quantum computational advantages for financial trading decision-making, presenting a solution for investors to optimize trading strategies. Through rigorous literature review, framework development, and comprehensive empirical evaluation, our work marks a substantial advancement in quantum finance.
The empirical results confirm the QADQN framework's robust performance, significantly surpassing standard frameworks and the buy-and-hold strategy in terms of total returns and risk-adjusted metrics during both test periods. This performance highlights the framework's effectiveness in adapting to diverse market scenarios and its capability to manage downside risks, evidenced by favorable Sharpe and Sortino ratios.
Despite these promising outcomes, our study recognizes the need for broader testing across varied market conditions and addresses the challenges related to quantum circuit scalability in practical applications. Future research should thus focus on enhancing the scalability of the QADQN framework, incorporating more complex quantum circuits, and expanding its application to different asset classes. Additionally, exploring quantum noise impacts and the potential for quantum error correction within the framework could further solidify the practical deployment of quantum-enhanced RL in financial markets.

\section*{Acknowledgment}
This work was supported in part by the NYUAD Center for Quantum and Topological Systems (CQTS), funded by Tamkeen under the NYUAD Research Institute grant CG008.

\bibliographystyle{IEEEtran}
\bibliography{refs}

\end{document}